# The MammoGrid Project Grids Architecture

Richard McClatchey, David Manset*, Tamas Hauer, Florida Estrella*, Pablo Saiz, Dmitri Rogulin*
*Centre for Complex Co-operative Systems, University West of England, Bristol BS16 1QY, UK*

Predrag Buncic
*Institut fur Kemphysik, August-Euler-Str.6, D-60486 Frankfurt, Germany*

*On behalf of the MAMMOGRID Consortium (CERN, Mirada-Solutions, Univs of Oxford, Sassari, Pisa, the West of England and the University Hospitals of Cambridge and Udine)*

\* Also ETT Division, CERN 1211, Geneva 23, Switzerland

The aim of the recently EU-funded MammoGrid project is, in the light of emerging Grid technology, to develop a European-wide database of mammograms that will be used to develop a set of important healthcare applications and investigate the potential of this Grid to support effective co-working between healthcare professionals throughout the EU. The MammoGrid consortium intends to use a Grid model to enable distributed computing that spans national borders. This Grid infrastructure will be used for deploying novel algorithms as software directly developed or enhanced within the project. Using the MammoGrid clinicians will be able to harness the use of massive amounts of medical image data to perform epidemiological studies, advanced image processing, radiographic education and ultimately, tele-diagnosis over communities of medical 'virtual organisations' .This is achieved through the use of Grid-compliant services [1] for managing (versions of) massively distributed files of mammograms, for handling the distributed execution of mammograms analysis software, for the development of Grid-aware algorithms and for the sharing of resources between multiple collaborating medical centres. All this is delivered via a novel software and hardware information infrastructure that, in addition guarantees the integrity and security of the medical data. The MammoGrid implementation is based on AliEn, a Grid framework developed by the ALICE Collaboration. AliEn provides a virtual file catalogue that allows transparent access to distributed data-sets and provides top to bottom implementation of a lightweight Grid applicable to cases when handling of a large number of files is required. This paper details the architecture that will be implemented by the MammoGrid project.

## 1. THE MAMMOGRID PROJECT

With the advent of the information age in radiology clinicians are being presented with analysis opportunities hitherto unforeseen, both in terms of data volumes and in data interpretation. Grids computing promises to resolve many of the difficulties in facilitating medical image analysis to allow clinicians to collaborate without having to co-locate. The EU-funded MammoGrid project [2] aims to investigate the feasibility of developing a Grid-enabled European database of mammograms so that a set of important healthcare applications using this database can be enabled and the potential of the Grids can be harnessed to support co-working between healthcare professionals across the EU.

Among the aims of having a Grids-enabled European-wide MammoGrid database are :

1. To evaluate current grid technologies and determine the requirements for Grid-compliance in a pan-European mammography database.
2. To implement the MammoGrid database, using novel Grid-compliant and federated-database technologies that will provide improved access to distributed data and will allow rapid deployment of software packages to operate on locally stored information.
3. To deploy enhanced versions of a standardization system (SMF [3]) that enables comparison of mammograms in terms of intrinsic tissue properties independently of scanner settings, and to explore its place in the context of medical image formats (such as Digital Imaging and Communications in Medicine, DICOM [4]).
4. To develop software tools to automatically extract image information that can be used to perform quality controls on the acquisition process of participating centres (e.g. average brightness, contrast).
5. To develop software tools to automatically extract tissue information that can be used to perform clinical studies (e.g. breast density, presence, number and location of micro-calcifications) in order to increase the performance of breast cancer screening programs [5].
6. To use the annotated information and the images in the database to benchmark the performance of the software described in points 3, 4 and 5.
7. To exploit the MammoGrid database and the algorithms to propose initial pan-European quality controls on mammogram acquisition and ultimately to provide a benchmarking system to third party algorithms.

The MammoGrid project concentrates on applying emerging Grid technology rather on developing it and plans to implement a lightweight (but fully functional) Grids technology and study its usage in hospitals. It will deliver a prototype federated database of mammograms in hospitals in the UK and Italy, in order to provide rapid feedback from the hospital community and inform the next generation of HealthGrids developments.

The MammoGrid information infrastructure which federates multiple mammogram databases, will enable clinicia[n]s to develop common, collaborative and co-operative approaches to the analysis of mammographic data. This paper outlines the MammoGrid Grid architecture for large-scale distributed mammogram analysis.

## 2. AN INFORMATION INFRASTRUCTURE

Grid technologies developed in the field of high energy physics is one possible answer to the MammoGrid challenge of storing, organizing and handling large data volumes and the need for accessing the data across geographically separated hospital sites. The MammoGrid project shares many of the requirements of high energy physics thus it is natural to investigate the feasibility of addressing those requirements with a solution.

Utilizing a Grid infrastructure for maintaining large numbers of medical images originating from multiple sources has a number of advantages. The volume of data – high number of cases and large (>10MB) image files – collected at and accessed from different locations already calls for a distributed data storage (with effective data-mining capabilities for image-related metadata). Together with the requirement of performing computationally expensive tasks (image analysis algorithms, simulations, quality control procedures) a grid-like solution indeed appears to be a justified design decision.

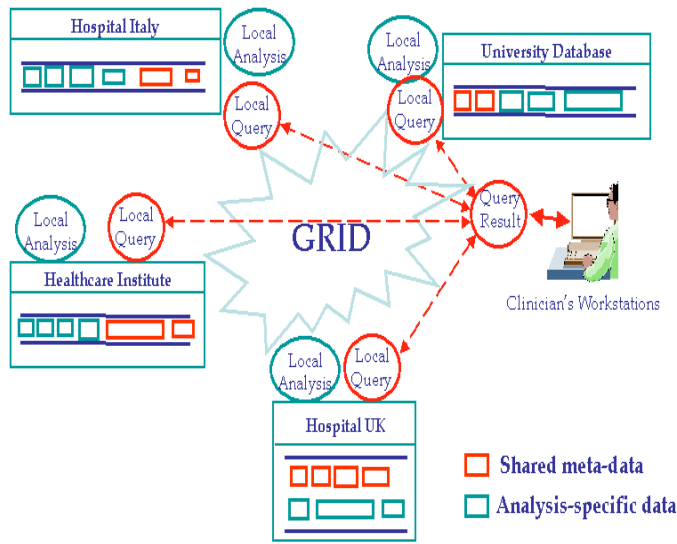

Figure 1 : Federated System Solution

Our proposed architecture shown in figure 1 is a federated system with the grid providing the necessary middleware for data distribution and transparency.
The information infrastructure for the MammoGrid project will be based on the philosophies behind two existing technologies:

- A lightweight Grid-compliant software package, AliEn [6], which employs a database-resident catalogue to co-ordinate the distribution and execution of jobs against massively distributed data files, and

- Reflective middleware, comprising high-level meta-data structures, based on the CRISTAL [7] software, which enables the location of data sets and the decomposition of complex queries into their constituent parts.

The MammoGrid project has adopted the Standard Mammogram Form (SMF [7]) as a common standard for mammogram analysis and storage across Europe and will develop SMF to cater for different acquisition protocols across Europe. MammoGrid will set the basis for a common database of mammography information comparable to other proposed systems in the US and the UK. SMF is not only the basis for standardisation of mammography, but it is also the key to quantitative assessment of breast images. MammoGrid will exploit this attribute in quality control and epidemiological studies. By participating in this project, consortium members aim to understand the implications of Grid technology in the deployment of its computer-aided design and computer aided diagnosis tools over the Internet.

Finally, MammoGrid is only a first stepping-stone in realising a pan-European database for mammography screening and analysis. Follow-up projects shall be sought which will include major European sites interested in running epidemiology research on the data available more generally across a generic healthcare grid. Similarly, the MammoGrid project can be extended to incorporate other disease groups and radiological applications.

### 2.1 Alien as a Grid Framework

AliEn (Alice Environment [6]) is a Grid framework developed to satisfy the needs of the Alice experiment at CERN LHC [8] (Large Hadron Collider) for large scale distributed computing for physics analysis. It is built on top of the latest Internet standards for information exchange and authentication (SOAP, SASL, PKI) and common Open Source components (such as Globus/GSI, OpenSSL, OpenLDAP, SOAPLite, MySQL,).

AliEn provides a virtual file catalogue that allows transparent access to distributed data-sets and provides top to bottom implementation of a lightweight Grid applicable to cases when handling of a large number of files is required (up to 2PB and $10^9$ files/year distributed on more than 20 locations worldwide in the case of the Alice experiment). At the same time, AliEn is meant to provide an insulation layer between different Grid implementations and provide a stable user and application interface to the community of Alice users during the expected lifetime of the experiment (more than 20 years).

### 2.2 CRISTAL as a Meta-Data Query Handler

CRISTAL [7] is a distributed scientific database system used in the construction and operation phases of HEP experiments at CERN. The CRISTAL project has studied the use of a description-driven approach using meta-data modelling techniques to manage the evolving data needs of a large community of scientists.

The advantage of following a description-driven design is that the definition of the domain-specific model (in this case the model of medical data) itself is captured in a computer-readable form and this definition may be interpreted dynamically by applications in order to achieve domain-specific goals. This approach has been shown to provide many powerful features such as scalability, system evolution, interoperability and reusability, aspects that are essential for future proofing medical information systems [9].

The current phase of CRISTAL research adopts an open architectural approach, based on a meta-model and a query facility to provide access to massively distributed data sets across a wide-area network and to produce an adaptable system capable of inter-operating with future systems and of supporting multiple user views onto a terabyte-sized database. It is thus ideally suited to being Grid-enabled as the basis of a MammoGrid query handler.

## 2.3 MammoGrid GRID Architecture

These two mature technologies have already proven their efficacy in delivering solutions for scalable database architectures at CERN and in handling distributed data volumes of the order of Terabytes. By combining these two technologies MammoGrid researchers will provide fresh insight into the mediation of queries across a geographically distributed database, will generate new approaches into the management of virtual organisations and will further the development of the next generation of Grid-resident information systems.

Since there is great flux and diversity in the many concurrent Grids projects being conducted around Europe, and since timely delivery of a Grid-compliant architecture for MammoGrid is crucial to the project, the implementation of an information infrastructure and Grid-compliance will be delivered in phases. In the first six months of the project (phase 1) a requirements analysis and hardware/software design study has been undertaken, together with a rigorous study of Grids software available from other concurrent projects. In addition to this, the CERN AliEn software is being installed and configured on a set of novel 'Grid-boxes', or secure hardware units, which will act as each hospital's single point of entry onto the MammoGrid. These units will then be configured and tested at CERN and Oxford, for later testing and integration with other Grid-boxes a the hospitals in Udine and Cambridge. As the MammoGrid project develops and new layers of Grids functionality become available, AliEn will facilitate the incorporation of new stable versions of Grids software ('plugging in' layers of software when available and suitably tested) in a manner that caters for controlled system evolution but provides a rapidly available lightweight but highly functional Grids architecture.

A limited but representative set of several hundred mammograms will be made available at CERN and Oxford to test out connectivity, accessibility and response. During months 7 to 18 the project will study and incorporate appropriate authentication protocols and application program interfaces (APIs) to this prototype mammographic database. This enables further tests to be undertaken, constituting the completion of phase 2 of the Information Infrastructure.

During the final phase of implementation and testing, lasting until the completion of the project, the meta-data structures required to resolve the clinicians' queries will be delivered using the meta-modelling facilities of the CRISTAL project. This will involve customizing a set of structures that will describe mammograms, their related medical annotations and the queries that can be issued against these data.

The meta-data structures will be stored in a MySQL [10] database at each node in the MammoGrid (e.g. at each hospital) and will provide information on the content and usage of (sets of) mammograms. The query handling tool will locally capture the elements of a clinician's query and will issue a query, using appropriate Grids software, against the meta-data structures held across multiple AliEn data centres in the distributed hospitals. At each location the queries will be resolved against the meta-data and the constituent sub-queries will be remotely executed against the mammogram databases. The selected set of matching mammograms will then be either analysed remotely or will be replicated back to the centre at which the clinician issued the query for subsequent local analysis, depending on the philosophy adopted in the underlying Grids software.

Concurrently with the meta-data study and periodically in the project lifespan, the architecture of the AliEn software will be reviewed (and replaced when appropriate Grids plug-ins become stable) in the light of new developments in other Grids software development efforts. A MammoGrid test-bed will enable a set of system tests to be carried out on the 18-month prototype mammogram database to ensure the security, accessibility and reliability of the test data samples and will provide a test harness under which the final MammoGrid database can be exercised by the identified medical use-cases with verification being in the hands of the Udine and Cambridge hospital user communities.

### 2.3.1 Overall Grids Architecture

Each site (hospital workstation) has direct, secure access to a dedicated Gridbox via the local area network. Each Gridbox can be seen as a "gate" to the Grid and is in charge of storing new Patient images / studies, updating the file catalogue and propagating the changes. The synchronous operation of the net of gridboxes ensures that the view of database at the workstations is up to date at every site. Gridboxes assume file transfers between them over the VPN network. The data sent through this network is anonymized and encrypted – an essential requirement for security and confidentiality.

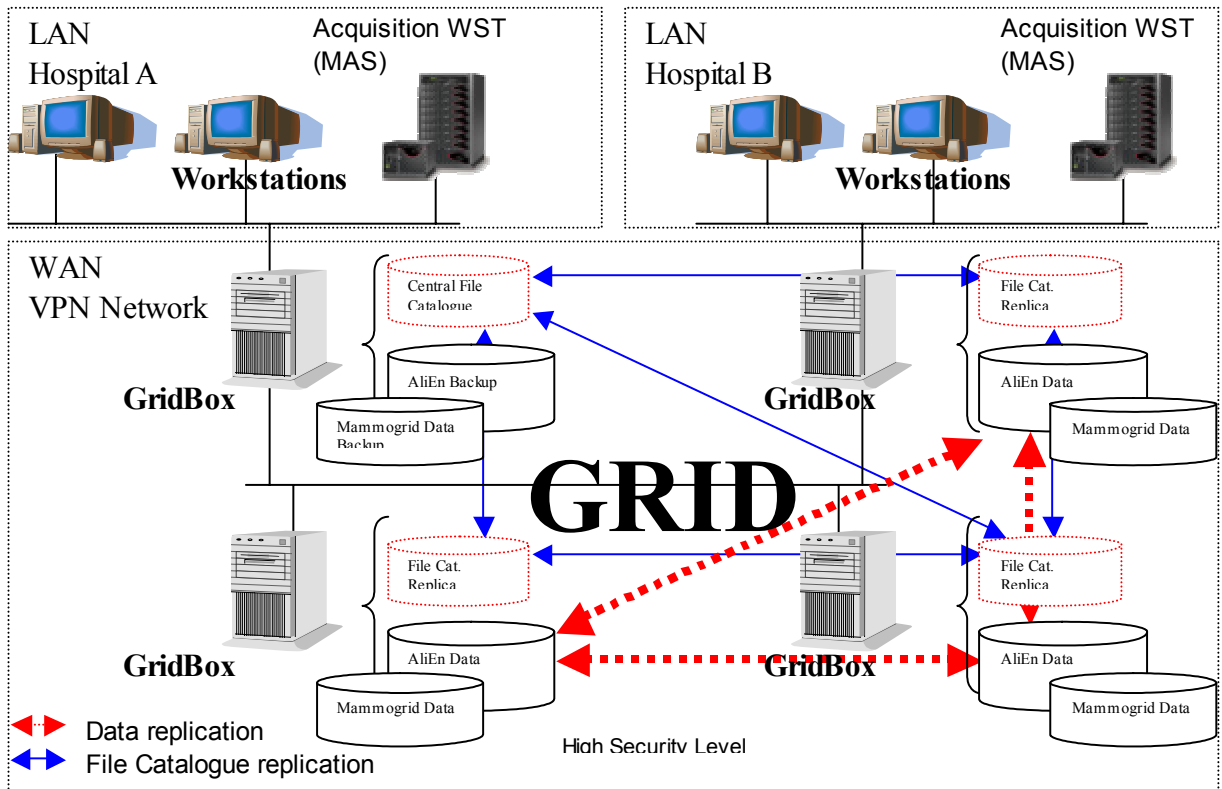

Figure 2 : Overall Grid Architecture

### 2.3.2 Local Architecture

While the GRID provides part of the essential features for the MammoGrid project, users need not (and should not) interact directly with grid functionalities. Rather, it is required that access to the mammogram database is exposed through a workstation client with a custom-made user interface. To that end, at every site (hospital), where a grid-box is deployed, at least one client machine is connected to the grid-box over a standardized communication/transfer protocol. The first prototype uses the default client of the MammoGrid project – a workstation developed by one of the partners (Mirada) – but the interface design aims to be general enough so that possibly other clients can be accommodated. This is achieved in three ways (1) the use of the industry-standard DICOM protocol for storing digital image and image-related data and (2) the use of standard-communication protocol model (SOAP and web services) for decentralized, distributed environment and (3) the use the W2C XML/XSD for data exchange format.

DICOM (Digital Imaging and Communications in Medicine) is a widely used standard for the interoperability of medical imaging equipments [4]. Its primary concern is the storage and sharing of digital images between medical imaging equipment and other systems. The MammoGrid project aims to conform to the DICOM standard in two ways. First, the digitized images should be imported and stored in the DICOM storage format (as dicom files), so that the full set of image- and patient-related metadata is readily available with the images, and that information exchange with other medical devices understanding the DICOM storage format is seamless. To further ensure the compatibility with DICOM conformant clients, it is required that the exchange of DICOM datasets should be done via the communication protocol – also defined by the standard. In this setup a client, or *Service Class User (SCU)* initiates a network connection with a server or *Service Class Provider (SCP)* and they exchange DICOM datasets over the established association protocol.

To facilitate DICOM compliance it is therefore required that the server side (grid-box) exposes a DICOM SCP which is capable of establishing an association with an SCU started by the client side (Mirada workstation). MammoGram X-rays should be transferred to the server for addition to the database via this association and similary requested image files are expected to be ready for download via the SCU/SCP pair.

The MammoGrid workstation should of course have communications capability which is richer than just transferring files back and forth. Besides being an image viewer component, it has to act as a user interface to the X-ray and patient information database, through which the users initiate database transactions, queries, updates and the like. For this purpose an API is defined so that the client can make and exchange structured data (as XML/XSD). This segment of the communication is done via an API deployed as a web service which can be accessed using SOAP. The simplified architecture of a site is shown on figure 3.

The DICOM and SOAP protocols, the W3C XML/XSD standard, a language-independent interface between the Mirada client and the Grid server, and a meta-data based data-model comprise the communication contract between core components of the Mammogrid system.

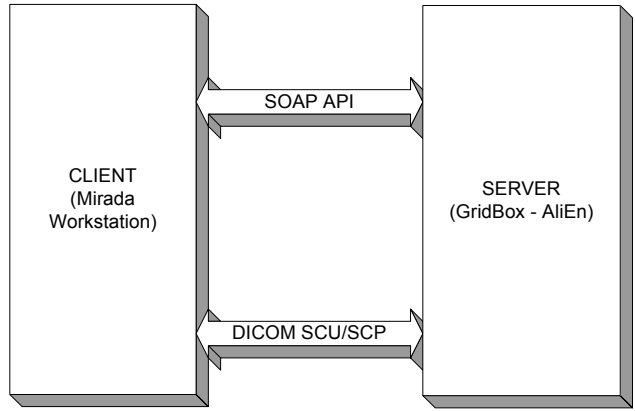

Figure 3 : Component View

These ensemble of standards and protocols bring about the potential for re-use (from the open-source computing and medical community) and for interoperability (with other standards-based system).

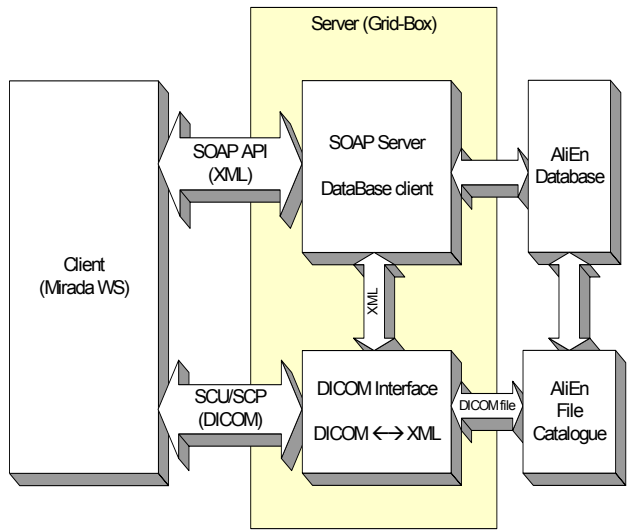

Figure 4 : Communication Diagram

Figure 4 illustrates the first deployed prototype. It shows the image and patient data exchange among the workstation, the Grid services and the data store. The workstation invokes Grid SOAP-based methods and transfers DICOM files down the DICOM file transfer protocol. The services of the Grid server transfer and exchange DICOM and XML files between them and to the data store. DICOM files are stored immutable, ie as files. In addition meta-data are stored separately to facilitate clinical queries and studies within and among files.

## 3. CONCLUSIONS

The MammoGrid project aims to investigate the feasibility of developing a European database of mammograms, accessed using emerging Grids software, so that a set of important healthcare applications using this database can be enabled and the potential of the Grids can be harnessed to support co-working between healthcare professionals across the EU. The main output of the 3-year MammoGrid project, launched in September 2002, is a Grid-enabled software platform (called the MammoGrid Information Infrastructure) which federates multiple mammogram databases and will enable clinicians to develop new common, collaborative and co-operative approaches to the analysis of mammographic data.

The motivation for the use of Grids technology in distributed image analysis for diagnosis, quality control, education and collaborative research is clear – Grids provide the mechanism for the sharing of large amounts of geographically distributed data with appropriate security and authentication to cater for the confidential nature of patient information. The MammoGrid project concentrates on the isolation of suitable Grids-enabling software technologies that provide the functionality for clinicians to co-operate without co-locating.

The MammoGrid project is adopting a philosophy that concentrates on the application of existing Grids middleware rather than on developing new Grids software. It is aimed primarily at an end-user community of radiologists and clinicians and investigates how the use of a Grid-compliant infrastructure can assist the end-users to resolve important clinical problems in mammogram image analysis. The project has gathered detailed user requirements from the University hospitals in Cambridge and Udine and is evaluating available Grids software, prior to embarking on the delivery of a prototype federated database system of mammograms which will be used to ensure the clinicians' requirements can be fulfilled and to provide feedback into the use of Grids software for medical applications. The project is also actively contributing to the EU-wide HealthGrid initiative [11] and a follow-up FP6 project is planned post-MammoGrid which will apply techniques emerging from MammoGrid to other pressing medical image analysis domains.

## 4. FUTURE WORK

This paper has described Grid information infrastructure that defines the borders of MammoGrid concepts and of its first implementation.

The first implementation of this architecture has been deployed and tested. Currently the team is working to incorporate our security model and new Grid services, like the query handler based on AliEn technologies, in order to complete the futur implementation and provide much more sophisticated possibilities. Moreover, communications between Grid services will no longer be based on RPC (Remote Procedure Call) but rather on message passing method, i.e. document literal.

We also plan to explore OGSA (Open Grid Services Architecture) to define MammoGrid Grid services OGSA compliant and extend our security model to see in what measure Globus GSI service could enhance it.

## Acknowledgments

We are grateful to many colleagues for numerous discussions on the topics discussed here, in particular Davd C. Schottlander and Mirada research and development team.